\documentstyle[epsfig,preprint,aps]{revtex}
\begin{document}
\preprint{hep-th/0205175.}
\tightenlines
\title{QUANTUM OSCILLATORS
IN THE CANONICAL COHERENT STATES}
\author{R. de Lima Rodrigues$^{(a)}$\thanks{Permanent address:
Departamento de Ci\^encias
Exatas e da Natureza, Universidade Federal de Campina Grande,
Cajazeiras, PB -- 58.900-000 -- Brazil},
A. F. de Lima$^{(b)}$, K. de Ara\'ujo Ferreira$^{(b)}$  and
A. N. Vaidya$^{(c)}$\\
{}$^{(a)}$ Centro Brasileiro de Pesquisas F\'\i sicas,
Rua Dr. Xavier Sigaud, 150\\
Rio de Janeiro-RJ-22290-180, Brazil\\
{}$^{(b)}$Departamento de F\'\i sica,
Universidade Federal de Campina Grande \\
 Campina Grande, PB --58.109-970 -- Brazil \\
{}$^{(c)}$ Instituto de F\'{\i}sica,
Universidade Federal do Rio de Janeiro \\
Ilha do Fund\~ao, Rio de Janeiro, RJ - 21.945-970 - Brazil}

\maketitle

\begin{abstract}
The main characteristics of the quantum oscillator
coherent states including the two-particle Calogero interaction
are investigated. We show that these Calogero coherent states are
the  eigenstates  of
the second-order differential annihilation  operator  which  is
deduced via R-deformed Heisenberg algebra or
Wigner-Heisenberg algebraic  technique and correspond exactly
to the pure uncharged-bosonic states. They possess the important properties
of non-orthogonality and
completeness.
The minimum uncertainty relation for the 
Calogero interaction coherent states is investigated.
New sets of Wigner oscillator even and odd coherent states are pointed out.

\end{abstract}

\vspace{0.5cm}

PACS numbers: 03.65.Fd, 03.65.Ge, 02.30.Tb
\pacs
\newpage

\section{INTRODUCTION \protect\\}
\label{sec:level1}

In the beginning of the sixties, the coherent states were investigated via
three definitions, viz., states of minimal uncertainty, eigenstates of the
annihilation operator and as being the states obtained by application of
the displacement operator on the ground state \cite{G}. In Ref. \cite{G} it
has been shown that these three definitions are equivalent for the simple
harmonic oscillator. Due to the fact that the energy spectrum of a particle
in a potential with centripetal barrier

\begin{equation}
\label{SP}
V(x) = {1\over 2} m\omega^2x^{2} + {1\over 2} gx^{-2},
\quad g=\lambda(\lambda +1), \quad -\infty  \le  x \le  \infty,
\end{equation}
is equally spaced like that of  the  simple  harmonic  oscillator, this
one-dimensional (1D) system is called an "isotonic oscillator" or
two-particle Calogero interaction \cite{GK}. In 3-dimensional space this
type of potential was first introduced by Davidson long ago \cite{Da}.
Considered first  by  Weissman  and Jortner \cite{WJ} in the context of
Gaussian wave functions, the  dynamics and  the energies of a coherent
states  for  the 1D isotonic oscillator  were studied.  Elsewhere  Nieto
and  Simmons Jr.  found the minimum-uncertainty coherent states (MUCS) and
discussed  various properties of this system, and   have also shown that
the three definitions of coherent states are equivalent for the simple
harmonic oscillator \cite{NS}. A  year later it was shown by Gutshick,
Nieto and Simmons Jr. \cite{GNS} that  the MUCS provide  us with a better
aproximation to the classical  motion  than do the Gaussians. In another
work Nieto \cite{N1} has shown  the mathematical and physical connection of
the charged-boson coherent  states \cite{BDB} with the MUCS. The canonical
coherent states for the Wigner generalized oscillator in the Schr\"odinger
representation were constructed by Sharma, Mehta and Sudarshan \cite{SMS}
and the representations and properties of para-bose oscillator operators
were investigated in a Schr\"odinger description \cite{SMMS}.

On the other hand, Leinaas and Myrheim \cite{LM88} have investigated the
relation bettween the
fractal  in 1+1 dimension and Calogero interaction,
and Fernandez {\it et al.} have investigated
the coherent states for SUSY partners of the oscillator \cite{fernandez99}.

In this work, we construct what we call canonical
coherent states (CCS) \cite{KS}, which are defined as
the eigenstates of the annihilation  operator
$B^{-}(\lambda)$
of the Calogero interaction Hamiltonian.
Such annihilation  operators are second-order differential ladder
operators \cite{D} and can be derived via the R-deformed
Heisenberg algebra or Wigner-Heisenberg algebraic
technique \cite{WH} which was recently
super-realized for the SUSY isotonic oscillator
\cite{JR1,Mik00}. The WH algebra has also been investigated for the
three-dimensional non-canonical oscillator to generate a representation
of the  orthosympletic Lie superalgebra $osp(3/2)$ \cite{PS}.
The coherent states of $SU(\ell, 1)$ groups  have been explicitly constructed
as orbits in some irreducible representations \cite{Git}.

The motion of the peaks of the wavefunctions for the coherent states of the
two-particle Calogero-Sutherland model were compared with the classical
trajectory \cite{AC}.
According to Calogero \cite{GK} the energy spectrum of the potential
(\ref{SP}) and $N$ bosons or fermions interacting are different by a
energy shift proportional to $\lambda=-\nu.$ Using an operator formulation
Brink {\it et al.} found all $N$-particle wave eigenfunctions
and extended the approach to the supersymmetric Calogero model
 and Heisenberg algebra in the simple case of two particles
\cite{FM}.
The observable for the two-anyon problem, satisfy the same algebra as
the observable in two-body Calogero problem.

Let us here point out the interesting connection between the
mesoscopic effects and
Calogero interaction  for a Coulomb gas under a new universality
in spectra of this chaotic system, which is described by a random
matrix theory \cite{Efetov}.

Another approach is the application of the time-dependent
parameters in the potential (\ref{SP}) in many quantum-mechanical effects.
For instance, Pedrosa {\it et al.} have used the Lewis-Riesenfeld invariant
method and a unitary transformation to obtain the exact Schr\"odinger
wave functions for a time-dependent harmonic oscillator with and without
an inverse quadratic potential \cite{Inacio97}.

Recently, Witten's supersymmetry formulation for
Hamiltonian systems \cite{Sukat95} has
been extended to a system of annihilation operator eigenvalue equations
associated with the supersymmetric unidimensional oscillator
and supersymmetric isotonic oscillator (singular potential),
which define supersymmetric
canonical coherent states
containing mixtures of both pure bosonic and pure fermionic
counterparts \cite{JRV99}. The breaking of supersymmetry due
to singular potentials
in supersymmetric quantum mechanics given by Eq. (\ref{SP})
has been recently investigated
\cite{Das00}.
In \cite{Mik00}, we see that the main result was
the observation of the intimate relationship between
the generalized statistics  and supersymmetry via R-deformed
Heisenberg algebra:
it was shown that the supersymmetry can be realized in purely parabosonic
systems (it is realized there in linear or nonlinear form depending on the
order of the paraboson).
This application has been considered for the
deformed Virasoro algebra so that
a representation of the modified Virasoro
algebra has been found \cite{elr01}.

Let us now point out that the R-deformed Heisenberg (or Wigner-Heisenberg)
algebra is given by following (anti-)commutation relations ($[A,B]_+\equiv
AB+BA$ and $[A,B]_-\equiv AB-BA):$

\begin{equation}
\label{RH}
H=\frac 12 [a^-, a^+]_+, \quad
[H, a^{\pm}]_-=\pm a^{\pm}, \quad
 [a^-, a^+]_-=1+\nu R, \quad
[R, a^{\pm}]_+=0,\quad R^2=1,
\end{equation}
where $\nu$ is a real constant associated to the Wigner parameter
\cite{JR1}. Note that when $\nu=0$ we have the
standard Heisenberg algebra.
The generalized quantum condition given in Eq. (\ref{RH}) has been
found relevant  in the context of integrable models \cite{Vasiliev91}.
Furthermore, this algebra was used for solving the energy
eigenvalue and eigenfunctions
of the Calogero interaction, in the context of one-dimensional
many-body integrable systems, in terms of a new set of phase
space variables involving exchanged operators \cite{Poly92,macfa93}.
Recently it has been employed for bosonization of supersymmetry
in quantum mechanics, and has also been demonstrated that
finite-dimensional representations are representations of the
deformed parafermionic algebra with internal
$Z_2-$ grading structure \cite{Mik97}.

Recently, the coherent states \`a la Klauder-Perelomov for a particle
moving in the P\"oschl-Teller potential of the trigonometric type have been
built up \cite{kinani}.

In this work,  we display some graphs showing the behavior of the
minimum uncertainty for a particular set of Wigner oscillator CCS.
This present work is organized as follows. In Sec. II, we start by
summarizing the R-deformed Heisenberg algebra or Heisenberg  algebraic
technique  for the  Wigner isotonic
oscillator \cite{JR1,Mik00}. While Jayaraman and Rodrigues,
in Ref. \cite{JR1}, adopt a super-realization of the
R-deformed Heisenberg algebra
as effective spectral resolution for the two-particle Calogero
interaction, in Ref. \cite{Mik00}, using the same super-realization,
Plyushchay
showed the various aspects of the R-deformed Heisenberg algebra.
In \cite{Mik00}, it was also shown that the nonlinear supersymmetry can be
realized
also at the classical level via the appropriate simple modification of
the model corresponding to the Witten supersymmetric quantum mechanics,
and it was noted that the quantization results in a generic case in the
quantum anomaly.

An generalization of the Heisenberg algebra which is written in terms of a
functional of one generator of the algebra has been analyzed by Curado and
Rego-Monteiro \cite{Reigo01}. The creation and annihilation operators of
correlated fermion pairs, in simple many body systems, satisfy a deformed
Heisenberg algebra that can be approximated by $q$ oscillators
\cite{Bonatsos}. In \cite{valdir}, a possility of extending the q-deformed
Heisenberg algebra to build a quantum field theory having fields that
produce at any space-time point particles satisfying the same algebra.
Also, Arik-Kili\c{c} have extended the $SU_q(2)$ algebra and
 the coherent states have been investigated \cite{arik01}.
The q-coherent states have been investigated for q-algebra related with
shape invariance condition by Fukui and Balantekin {\it et al.} \cite{balan99}.

  Using the Gazeau-Klauder approach \cite{GK96} for coherent states
  associated with quantum systems,  Antoine {\it et al.} \cite{antoine01}
  have analyzed the spatial and temporal features
  of the coherent states associated to the infinite square-well
and P\"oschl-Teller potentials. Also, Daoud-Hussin have found
new general sets of coherent states and the 
quantum optics Jaymes-Cummings model \cite{Daoud01}.
  Moreover, Popov has constructed and
investigated the pseudoharmonic oscillator in the Barut-Girardello coherent
states and photon-added Barut-Girardello coherent states \cite{popov02}.

 In Sec. III, we define and
build without supersymmetry the two-particle Calogero interaction canonical
coherent states as the eigenstates of the second-order differential
annihilation operator ($B^{-})$. In Sec. IV, we discuss the MUCS from
Wigner first-order differential ladder operators and Calogero interaction
ladder operators. In this Section, new even and odd canonical coherent
states are pointed out. Sec. V contains the conclusions.

\section{The Super Wigner-Heisenberg Algebra}

For convenience we choose units so that $\hbar=\omega =m=1.$ Thus,
for a super-realization of the R-deformed Heisenberg algebra (\ref{RH}),
the system governed by the potential
(1)  becomes  identical  to  the  potential of the
bosonic  sector  of  the  Wigner
Hamiltonian. The 1D Wigner oscillator Hamiltonian in terms of
the Pauli's matrices ($\sigma_i,$ i=1,2,3) is given by

\begin{eqnarray}
H(\lambda +1) &=& {1\over 2}\left\{- {d^{2}\over dx^{2}} + x^{2} +
{1\over x^{2}} (\lambda +1)[(\lambda +1)\sigma_{3} - 1]\sigma_{3}\right\}
\nonumber\\
&=& \left(
\begin{array}{cc}
H_{-}(\lambda )&0\\
0&H_{+}(\lambda )=H_{-}(\lambda +1)
\end{array}\right),
\end{eqnarray}
where the even and odd sector Hamiltonians are respectively given by

\begin{equation}
\label{ich}
 H_{-}(\lambda ) = {1\over 2}\left\{ - {d^{2}\over dx^{2}} +
x^{2} + {1\over x^2} \lambda (\lambda +1)\right\}
\end{equation}
and
\begin{equation}
H_{+}(\lambda ) = {1\over 2} \left\{- {d^{2}\over dx^{2}} + x^{2} +
{1\over x^{2}} (\lambda +1)(\lambda +2)\right\}= H_{-}(\lambda +1).
\end{equation}
The even sector is the Hamiltonian of the oscillator with barrier.

Note that the Wigner oscillator ladder operators

\begin{equation}
\label{loa}
 a^{\pm} = \frac{1}{\sqrt {2}} (\pm i\hat {p}_x - \hat x)
\end{equation}
of the R-deformed Heisenberg algebra may be written in terms of the
super-realization of the position and momentum operators viz., $\hat
x=x\sigma_1$ and $\hat{p}_x=-i\sigma_1\frac{d}{dx}+\frac{\nu}{2x}\sigma_2,$
satisfy the general quantum rule $[\hat {x}, \hat {p}_x]_- = i (1 + \nu\hat
{R}),$ where $\nu=2(\lambda +1).$ Thus, in this representation the
reflection operator becomes $R=\sigma_3,$ where $\sigma_3$ is the diagonal
Pauli matrix.

Thus, from the super-realized first order ladder operators  given by
\cite{JR1,Mik00}

\begin{equation}
a^{\pm }(\lambda +1) = {1\over \sqrt{2}}\left\{\pm {d\over dx} \pm
{(\lambda +1)\over x}\sigma_{3} - x\right\}\sigma_{1},
\end{equation}
the Wigner Hamiltonian becomes
\begin{equation}
\label{E6}
H(\lambda +1) = {1\over 2}\left[a^{+}(\lambda +1), a^{-}(\lambda +1)\right]_{+}
\end{equation}
and the Wigner-Heisenberg algebra ladder relations are readily obtained as

\begin{equation}
\label{E7}
\left[H(\lambda +1), a^{\pm }(\lambda +1)\right]_{-} =
\pm a^{\pm }(\lambda +1).
\end{equation}
Equations (\ref{E6}) and (\ref{E7}) together with the commutation relation
\begin{equation}
\label{E8}
\left[a^{-}(\lambda +1), a^{+}(\lambda +1)\right]_{-}=
1 + 2(\lambda +1)\sigma_{3}
\end{equation}
constitute the R-deformed Heisenberg algebra.

The Wigner eigenfunctions that generate
the eigenspace associated with even(odd) $\sigma_{3}$-parity for even(odd)
quanta $n=2m (n=2m+1)$ are given by
\begin{equation}
\mid n=2m, \lambda +1> = \left(\begin{array}{cc} \mid m,\lambda > \\ 0
\end{array}\right),\quad
\mid n=2m+1, \lambda> =
\left(\begin{array}{cc} 0 \\ \mid m,\lambda >
\end{array}\right)
\end{equation}
and satisfy the following eigenvalue equation

\begin{equation}
H(\lambda +1)\mid n, \lambda +1> = E^{(n)}\mid n, \lambda +1> ,
\end{equation}
where the non-degenerate energy eigenvalues are obtained by the
application of the raising operator on the ground eigenstate and are given by
\begin{equation}
E^{(n)} = \lambda  + {3\over 2} + n, \quad n=0,1,2,\ldots .
\end{equation}

For the oscillator with barrier the energy  eigenvectors  satisfy  the
following equations
\begin{equation}
H_{-}(\lambda )\mid m, \lambda > = E^{(m)}_{-}\mid m, \lambda >,
\end{equation}
where the eigenvalues are exactly constructed via R-deformed Heisenberg
algebra ladder relations
and are given by \cite{JR1}

\begin{equation}
E^{(m)}_{-}=
\lambda + {3\over 2} + 2m ,\qquad m=0,1,2,\ldots .
\end{equation}
Also from the Wigner-Heisenberg algebra  we  obtain  the  second-order
differential raising and lowering operators for the energy spectrum
of the 1D oscillator with barrier,viz.,
on ${1\over 2}(1 + \sigma_3)$ projection, the
R-deformed Heisenberg algebra representations
decouple, $\left[H(\lambda +1), {a^{\pm }}^2(\lambda +1)\right]_{-} =
\pm 2{a^{\pm }}^2(\lambda +1).$ Indeed, the left hand side leads us

$$
{1\over 2}(1+\sigma_3)
[H(\lambda +1),
{a^{\pm }}^2(\lambda +1)]_{-} =
\left(
\begin{array}{cc}
[H_{-}(\lambda), B^{\pm}(\lambda)]_-&0\\
0&0
\end{array}\right)
$$
and the right hand side becomes

$$
{1\over 2}(1+\sigma_3){a^{\pm }}^2(\lambda +1)=
\left(
\begin{array}{cc}
B^{\pm}(\lambda)&0\\
0&0
\end{array}\right),
$$
where

\begin{equation}
\label{B-}
B^{-}(\lambda) = {1\over 2}\left \{{d^{2}\over dx^2} +
2x{d\over dx} + x^{2} - {\lambda (\lambda +1)\over x^{2}} + 1\right\}
\end{equation}
and

\begin{equation}
\label{B+}
B^{+}(\lambda) = {1\over 2}\left\{{d^{2}\over dx^2}-
2x{d\over dx} + x^{2} - {\lambda (\lambda +1)\over x^{2}} - 1\right\}.
\end{equation}
Thus, these ladder operators obey the following commutation relations:
\begin{eqnarray}
[B^-(\lambda), B^ +(\lambda)]_-&=&4H_-(\lambda) \nonumber\\
\left[H_-, B^{\pm }(\lambda)\right]_- &=& \pm 2B^{\pm }(\lambda) .
\end{eqnarray}
Hence, the quadratic operators $B^{\pm }(\lambda)$ acting on
the orthonormal basis of eigenstates of $H_-(\lambda),$
$\{ \mid m, \lambda > \}$ where $m=0,1,2, \cdots$
have the effect of  raising  or  lowering  the
quanta by two units so that we can write

\begin{equation}
\label{EA}
B^{-}(\lambda)\mid m, \lambda > = \sqrt{2m(2m+2\lambda +1)}
\mid m-1, \lambda >
\end{equation}
and
\begin{equation}
\label{EL}
B^{+}(\lambda)\mid m, \lambda > = \sqrt{2(m+1)(2m+2\lambda +3)}
\mid m+1, \lambda >
\end{equation}
giving

\begin{equation}
\label{E19}
\mid m, \lambda > = 2^{-m}\left\{{\Gamma (\lambda +3/2)\over m!
\Gamma (\lambda +m+3/2)}\right\}^{1/2}\left\{B^{+}(\lambda)\right\}^{m}
\mid 0, \lambda >,
\end{equation}
where $\Gamma (x)$ is the ordinary Gamma Function.
Note that $B^{\pm }(\lambda)\mid m, \lambda >$ are
associated with the energy eigenvalues
$E^{(m\pm 1)}_- =\lambda + {3\over 2} + 2(m\pm 1),
\quad m=0, 1, 2, \ldots .$

Let us to conclude this section presenting a very simple question:
what is the
structure generated by the new operators pointed out in this
section from quantum oscillator? Note that the operators
$\pm \frac{i}{2}\left(a^{\pm}(\lambda +1)\right)^2$ and
$\frac 12 H(\lambda +1)$ can be chosen as
a basis for a realization of the $SO(2,1)\sim SU(1,1)\sim SL(2,{\bf R})$
Lie algebra.
 When  projected the $-\frac 12 \left(a^{\pm}\right)^2$ operators in the even
sector we obtain that $-\frac 12 B^{\pm}$ and together with $\frac 12 H_{-}$
generate once again the Lie algebra $SU(1,1).$ 

Consequently, for Calogero interaction the resultanting Lie algebra is
$SU(1,1):$

\begin{equation}
[ K_0, K_1]_-=iK_2, \quad 
[ K_1, K_2]_-=-iK_0, \quad
[ K_2, K_0]_-=-iK_1.
\end{equation}
Indeed, from (\ref{ich}),
(\ref{B-}) and (\ref{B+}) we obtain
$K_0=\frac{H_-}{2}, K_1=-\frac 14 (B^{-} +B^{+})$ and 
$K_2=-\frac i4 (B^- -B^+).$
Therefore one can generate the generalized  coherent  states according to
Perelomov \cite{Perelomov72,Fujii01}.

\section{Calogero Interaction Canonical Coherent States}

Now, we define the
Calogero interaction canonical coherent states, $\mid\alpha, \lambda>$,  as
the eigenkets of the annihilation
operator $B^{-}(\lambda)$,

\begin{equation}
\label{EEA}
B^{-}(\lambda)\mid\alpha, \lambda> = \alpha\mid\alpha,\lambda>,
\end{equation}
where the eigenvalue $\alpha $ can be  any  complex  number.  Writing
\begin{equation}
\label{cics}
\mid\alpha, \lambda> =\sum^{\infty }_{m=0} b_{m}\mid m, \lambda>
\end{equation}
we  obtain  a  recursion
relation for the coefficients $b_{m}$

\begin{equation}
b_{m} = {\alpha \over 2} \left\{m(m + \lambda +
\frac{1}{2})\right\}^{-\frac{1}{2}}b_{m-1}=\frac{(\frac{\alpha}{2})^{m}}{\{m!
\Gamma (m+\lambda +\frac{3}{2})\}^{\frac{1}{2}}}b_0
\end{equation}
which provides us with the normalized canonical coherent states $(<\alpha,
\lambda \mid\alpha, \lambda >)$ for the Calogero interaction in the form

\begin{equation}
\label{ECCS}
\mid\alpha, \lambda > = \{ g(\mid\alpha\mid ) \}^{-\frac{1}{2}}
\sum^{\infty }_{m=0}\frac{(\frac{\alpha}{2})^{m}}{\{m! \Gamma (m+\lambda
+\frac{3}{2})\}^{\frac{1}{2}}} \mid m, \lambda>,
\end{equation}
where the normalization constant $b_0$ is given by

\begin{equation}
\label{ECON} b_0^2=g(\mid \alpha \mid ) = \left\{{2\over \mid \alpha \mid
}\right\}^ {(\lambda +\frac{1}{2})}I_{\lambda +\frac{1}{2}}(\mid \alpha
\mid )
\end{equation}
and $I_{\nu }(\mid \alpha \mid )$ is the modified Bessel function
of the  first  kind,
\begin{equation}
\label{EBE}
I_{\nu }(\mid \alpha \mid ) =
\sum^{\infty }_{m=0}\frac{\{{\mid \alpha \mid \over 2}\}^
{(2m+\nu )}}{m! \Gamma (m+\nu +1)}.
\end{equation}

The Calogero interaction CCS are normalized
however they are non-orthogonal since

\begin{equation}
\label{EN} <\xi ,\lambda \mid \alpha ,\lambda > = \left\{g(\mid \alpha \mid
) g(\mid \xi \mid )\right\}^{-\frac{1}{2}} g\left((\xi^{*}\alpha
)^{\frac{1}{2}}\right),
\end{equation}
which means that the CCS is an over-complete. The resolution of unity is
given by
\begin{equation}
\int\mid \alpha, \lambda ><\alpha, \lambda \mid
{1\over 2\pi } K_{\lambda +\frac{1}{2}}(\mid \alpha \mid )
I_{\lambda +\frac{1}{2}}(\mid \alpha \mid )
d^2\alpha =
\sum^{\infty }_{m=0}\mid m, \lambda><m, \lambda\mid  = {\bf 1},
\end{equation}
where $x= \mid \alpha \mid, z=1$ and $t=\sinh u$ with

\begin{eqnarray}
K_{\nu }(\mid \alpha \mid ) & =  &
2^{\nu }{\Gamma (\nu +\frac{1}{2})\over \mid \alpha \mid ^{\nu }\sqrt{\pi}}
 \int^{\infty}_0
{\cos (\mid \alpha \mid t)\over (t^{2} + 1)^{\nu +\frac{1}{2}}} dt
\nonumber\\
& \mbox{} = &
{2\Gamma (\nu +\frac{1}{2})\over \mid \alpha \mid\sqrt{\pi}}
\int^{\infty}_0
(\cosh u)^{-2\nu}\cosh (\mid \alpha \mid \sinh u) du,
\end{eqnarray}
which is a particular form of $K_{\nu }(zx)$ is the
modified Bessel function of the third kind \cite{GR}.

The completeness property here deduced is formally analogous to the
resolution of the identity for the isotonic oscillator minimum-uncertainty
coherent states \cite{N1}. However, one obtain this properties from our
operators deduced via the super-realization of the R-deformed Heisenberg.

\section{THE MINIMUM UNCERTAINTY COHERENT STATES}
\label{sec:level4}

Let us begin by making some remarks about the CCS and the minimum
uncertainty coherent states (MUCS) of Wigner oscillator. The CCS of Wigner
annihilation operator which satisfies the R-deformed Heisenberg algebra are
defined by

\begin{equation}
\label{WCS} a^-|\zeta>_W\, =\zeta |\zeta>_W,
\end{equation}
and can be written in terms of the Wigner oscillator eigenstates

\begin{equation}
\label{ESP} \mid\zeta>_W =\sum^{\infty }_{n=0} c_{n}\mid n, \lambda+1>
\end{equation}
where the eigenvalue $\zeta$ can be any complex number.

From the definition of the position $\hat x$ and momentum $\hat p,$ quantum
operators given by Eqs. (\ref{loa}) and (\ref{E8}) we have the following
commutation relation

\begin{equation}
[\hat {x}, \hat {p}_x]_- = i [1 + 2(\lambda+1)\sigma_3].
\end{equation}
Thus, there is a generalised uncertainty relation for  $\hat x$ and $\hat
p,$ given by

\begin{equation}
\Delta \hat x \Delta \hat p\geq\frac 12\mid 1+2(\lambda +1)<\sigma_3>\mid.
\end{equation}
 We show that the
product of uncertainties of the Wigner oscillator position $\hat x$ and
momentum $\hat p,$ in the CCS becomes minimal among those values which are
permissible by quantum mechanics, viz., $\Delta \hat x \Delta \hat
p=\mid\frac 12+(\lambda +1)<\sigma_3>\mid,$ where $<\sigma_3>$ is the
average value in the Wigner oscillator CCS. For instance note that when
$\lambda=-1$ the unidimensional oscillator MUCS  is re-obtained. A detailed
analysis of Wigner oscillator coherent states is in preparation.

To complete our analysis on coherent states for the  Calogero interaction,
we trace below the construction of minimum uncertainty coherent states. Now
let us consider new definitions of the position $\hat X$ and momentum $\hat
P$ so that

\begin{equation}
B^ {\mp}=\frac{1}{\sqrt{2}}(\mp i\hat P -\hat X)
\end{equation}
which leads us to the following commutation relation

\begin{equation}
[\hat X, \hat P]=4iH_-.
\end{equation}

In this case the minimum uncertainty states $\mid \alpha >_M$ with equal
dispersions for $\hat X$ and $\hat P$ are given by

\begin{equation}
\label{EEA2}
B^ -\mid \alpha>_M=\alpha \mid\alpha>_M, \quad
\alpha=-\frac{1}{\sqrt{2}}(<\hat X>+i<\hat P>)
\end{equation}
where $\mid\alpha, \lambda >_M$  is an eigenstate particular set
of $B^-$.

Therefore, using the two identities

\begin{eqnarray}
{\hat X}^2 &&= \frac{1}{2} \left((B^-)^2+(B^+)^2+2B^+B^-+4H_-\right),
\nonumber\\ {\hat P}^2 &&=-\frac{1}{2}
\left((B^-)^2+(B^+)^2-2B^+B^--4H_-\right)
\end{eqnarray}
we obtain the following expectation values in the CCS:

\begin{eqnarray}
\label{EVXP} <{\hat X}>&&= -\frac{1}{\sqrt 2} \left(\alpha^*+\alpha\right)=
-\sqrt{2}Re(\alpha), \nonumber\\ <{\hat P}>&&=\frac{i}{\sqrt 2}
\left(\alpha^*-\alpha\right)= \sqrt{2}Im(\alpha) \nonumber\\ <{\hat
X}^2>&&= 2[Re(\alpha)]^2+2<H_->
 \nonumber\\
<{\hat P}^2>&&=2[Im(\alpha)]^2+2<H_->,
\end{eqnarray}
where

\begin{equation}
\label{VEHOB} <H_->=<\alpha\mid
H_-\mid\alpha>_M=\mid\alpha\mid\frac{I_{\lambda-\frac 12}(\mid\alpha\mid)}
{I_{\lambda+\frac 12}(\mid\alpha\mid)} +\frac 12 -\lambda.
\end{equation}

The variances of position operator $((\Delta \hat X)^2=<{\hat X}^2>-<{\hat
X}>^2)$ and momentum operator $((\Delta \hat P)^2=<{\hat P}^2>-<{\hat
p}>^2)$ on the coherent states $ |\alpha>_{M}$ are identical. Next,
$I_{\lambda-\frac 12}(\mid\alpha\mid)$ is  the modified Bessel function of
the first kind given by Eq. (\ref{EBE}), for  $x>>1$ is given by

\begin{equation}
\label{bi} I_n(x)\sim \frac{e^x}{\sqrt{2\pi x}}\left(1-\frac{4n^2-1}{8x}+
\frac{(4n^2-1)(4n^2-3^2)}{2!(8x)^2}+\cdots\right).
\end{equation}
Thus, from Eq. (\ref{EVXP}) the minimum uncertainty relation for
$\mid\alpha\mid>>1,$
 $\lambda=\frac 12$ and $\lambda=0,$ respectively, becomes

\begin{eqnarray}
\label{RIMI}
\Delta \hat X \Delta \hat P &&=\mid\alpha\mid\nonumber\\
\Delta \hat X \Delta \hat P &&=\mid\alpha\mid +\frac 12.
\end{eqnarray}

Indeed, we find that the variances in new position and momentum satisfy

\begin{eqnarray}
\label{RIM} <(\Delta \hat X)^2>&&=2<H_->= <(\Delta \hat P)^2>\nonumber\\
\Delta \hat X \Delta \hat P&&=2\mid\alpha\mid\frac{I_{\lambda-\frac
12}(\mid\alpha\mid)} {I_{\lambda+\frac 12}(\mid\alpha\mid)} +1 -2\lambda.
\end{eqnarray}
Therefore, we show that a Calogero interaction in CCS leads us to minimum
uncertainty relation.
 In figures I, II and III we plot
$<{\hat X}^2>$  and $(\Delta \hat X)(\Delta \hat P),$ for two particular
values of $\lambda.$

\section{CONCLUDING REMARKS}
\label{sec:level5}

We have presented the canonical coherent states (CCS) associated with the
unidimensional harmonic oscillator plus a centripetal barrier (a Calogero
interaction \cite{GK,Perelomov72} with two particles for the relative
coordinate $x=x_1-x_2$ or isotonic oscillator \cite{D}), which preserve the
property of non-orthogonality. These CCS were deduced via R-deformed
Heisenberg (or Wigner-Heisenberg) algebra in non-relativistic quantum
mechanics. Although we have mainly treated the Calogero interaction CCS,
similar results can be adequately extracted for any physical D-dimensional
radial oscillator system by the Hermitian replacement of $ - i{d \over dx}
\rightarrow - i \left({d\over dr} + {D-1\over 2r}\right)$ and the Wigner
deformation parameter $\lambda + 1 \rightarrow \ell_D + {1\over 2} (D - 1)$
where $ \ell_D  (\ell_D = 0, 1, 2, ...) $ is the D-dimensional oscillator
angular momentum. In tridimensional space this Hermitian replacement left
us exactly to the potential first investigated  by Davidson in the
beginning of thirties \cite{Da}.

 Therefore we can construct new spherical coherent states for diatomic
molecules with Davidson interaction \cite{RB}, so that complete diatomic
molecule energy spectra and eigenfunctions can be deduced algebraically via
R-deformed Heisenberg algebra or Wigner-Heisenberg factorization method
\cite{JR1,Mik00}.

We also consider a succinct anylis of the construction of minimum
uncertainty coherent states (MUCS) for the Wigner oscillator position $\hat
x$ and momentum $\hat p,$ and a detailed anylis of minimum uncertainty
coherent states for the Calogero interaction.

Let us point out that a CCS $|z>_{\hbox{W}}$ is an eigenstate of the Wigner
annihilation operator according to Eq. (\ref{WCS}), it is possible to show
that the analogous of so called even and odd CCS $|z,\pm>$, which appear in
the coherent states for the usual harmonic oscillator and Quantum Optics for
uncharged quanta\cite{Gerry93} and charge quanta \cite{XL}, are eigenstates
of the operator $\left(a^-(\lambda+1)\right)^2$ (but not of
$a^-(\lambda+1)).$ A detailed analysis of the generation of even and odd
canonical coherent states via R-deformed Heisenberg algebra will be
published elsewhere.

Recently Alexanian {\it et al.} have built a star product associated 
with an
arbitrary two-dimensional Poisson structure using generalized coherent
states on the complex plane \cite{Alex01}. The coherent states for the
isotonic oscillator  has been considered in the coordinate representation
by Bagchi and Bhaumik \cite{BBD}. The correspondence between eigenvalue
Eqs. (26) in Ref. \cite{BBD} and our Eq. (\ref{EEA}) provided us
$z=\frac{\alpha}{\sqrt{2}}$, where $2z^2$ is the eigenvalue in Ref.
\cite{BBD}.

Following  \cite{FM,Efetov,Perelomov72} it is of interest to note
that defining

$$
A^{\pm}_j=\frac{1}{\sqrt 2}\left\{p_j\pm
i\left(\frac{\partial}{\partial x_j}W(x_j)\right)\right\}, \quad p_j
=-i\frac{d}{dx_j}
$$
we obtain in one dimension the factorized Hamiltonian of the Calogero
interaction for a many particle system with Davidson interactions

$$
H_-=\sum_jA^+_jA^-_j= \frac 12\sum_j\left\{p_j^2+
\left(\frac{\partial}{\partial x_j}W(x_j)\right)^2\right\}+
\frac 12\sum_{i,j}\frac{\partial^2 W(x_j)}{\partial x_i\partial x_j}.
$$
Thus, following \cite{Efetov}, under the context of mesoscopic physics,
in the case of $N$-boson or fermion Hamiltonian
we can obtain the relation between the Brownian motion and
Calogero model.

For instance in our case with two particles the choice
$
W(x)=\frac 12 x^2+(\lambda +1)\ell n (x)
$
gives us $H_-$ belonging to the even sector of
$H(\lambda +1).$
These aspects will be considered elsewhere in construction
of the supercoherent states \cite{JRV99} for diatomic molecules
\cite{RB}. However, note that in this work
the $A^{\pm}_j$ operators become

$$
A^{\pm}_j\rightarrow {\cal A}^{\pm}=
\frac{1}{\sqrt 2}\left(-i\frac{d}{dx}\pm i\frac{(\lambda +1)}{x}
\pm ix\right).
$$

Therefore, the present work opens a new route for future investigations on
the Calogero interaction coherent states, for instance let us point out
that the R-deformed Heisenberg (or Wigner-Heisenberg) algebra can be
applied for a complete spectral resolution of the complex Calogero model
with real energies \cite{ZT}, too. Finally, let us point out that one can
consider an analysis of the Calogero interaction coherent states as
reported in the works of references \cite{baha00,Siva,jellal02}.

\vspace{2cm}

\centerline{\bf Acknowledgments}

This research was supported in part by CNPq (Brazilian Research Agency).
RLR wish to thanks the staff of the CBPF and DCEN-CFP-UFPB for the
facilities. Thanks are also due to J. A. Helayel Neto for the kind of his
hospitality of RLR at his post-doctoral traineeships in CBPF-MCT. The
authors would also like to thank M. Plyushchay, A. M. S. Macedo, I. A.
Pedrosa, M. A. Rego-Monteiro, D. J. Fernandez, Polychronakos, M. C\^andido
Ribeiro and A. Aleixos for the kind interest in pointing out relevant
references on the subject of this paper.


\unitlength=1cm
\begin{figure}[tbp]
\centering
\begin{picture}(10,1)
\epsfig{file=RIIC101.eps,width=10cm,height=08cm,angle=-90}
\end{picture}
\vspace{7.5cm}
\caption{The minimum uncertainty relation for the Calogero system
coherent states given by Eqs.(\ref{VEHOB}) and (\ref{RIM}),
for $\lambda=\frac 12$}
\end{figure}


\unitlength=1cm
\begin{figure}[tbp]
\centering
\begin{picture}(10,1)
\epsfig{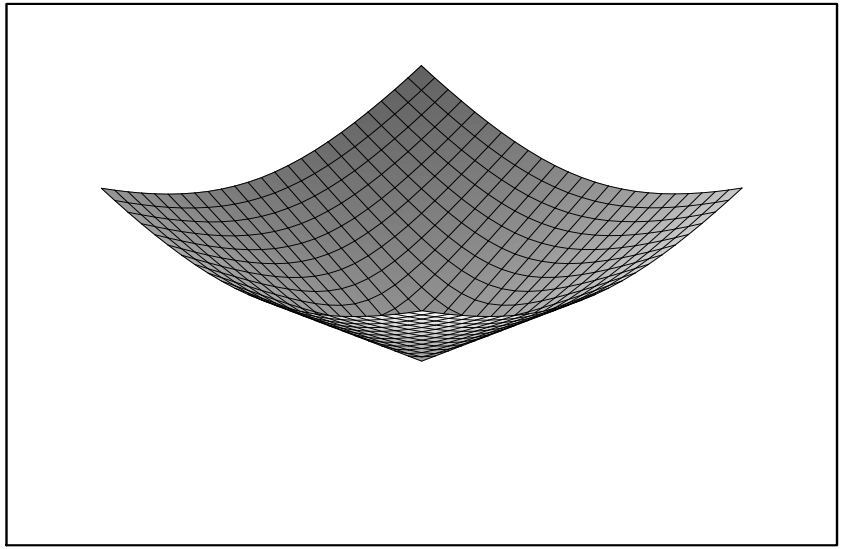}
\end{picture}
\vspace{7.5cm}
\caption{The minimum uncertainty relation for the Calogero system
coherent states for $\lambda=\frac 12$ and $\mid\alpha\mid>>1$ is
given by Eq. (\ref{RIMI}).
}
\end{figure}

\newpage


\unitlength=1cm
\begin{figure}[tbp]
\centering
\begin{picture}(10,1)
\epsfig{file=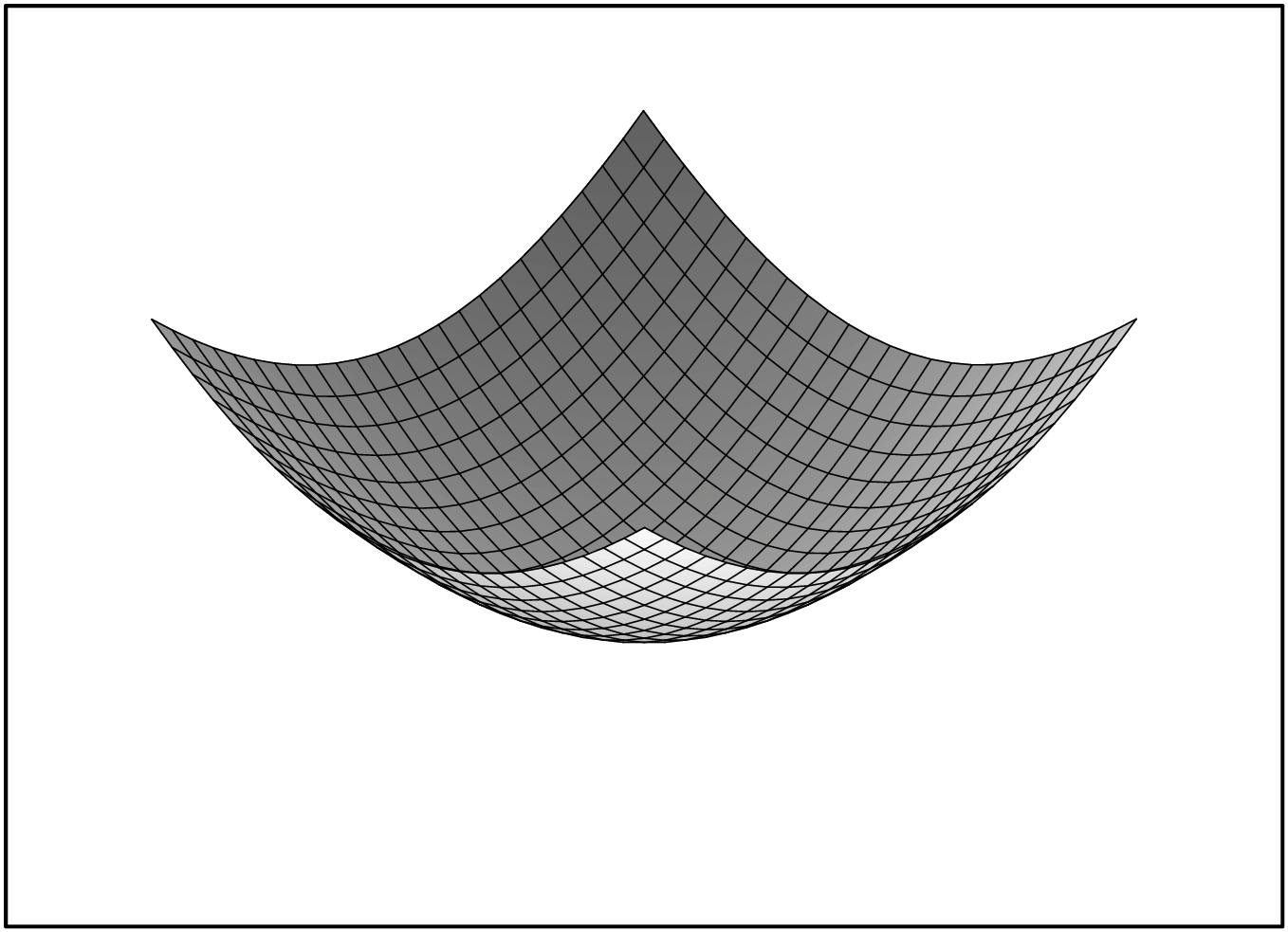,width=10cm,height=08cm,angle=-90}
\end{picture}
\vspace{7.5cm}
\caption{The minimum uncertainty relation for the Calogero system
coherent states given by Eqs.(\ref{VEHOB}) and (\ref{RIM}),
for $\lambda=10.0$}
\end{figure}

\end{document}